\documentstyle[epsf,12pt,a4]{article}
\newcommand{\be}{\begin{equation}}
\newcommand{\ba}{\begin{eqnarray}}
\newcommand{\ee}{\end{equation}}
\newcommand{\ea}{\end{eqnarray}}
\newcommand{\nn}{\nonumber}

\newcommand{\GeV}{\;\mbox{GeV}}
\newcommand{\MeV}{\;\mbox{MeV}}
\newcommand{\eV}{\;\mbox{eV}}

\newcommand{\secs}{\;\mbox{s}}
\newcommand{\simlt}{\stackrel{<}{{}_\sim}}
\newcommand{\simgt}{\stackrel{>}{{}_\sim}}
\newcommand{\ol}{\overline}

\newcounter{currequation}
\begin{document}
\title{
Uses of a small field value which falls from a metastable maximum over cosmological times
}
\author{Saul Barshay\footnote{barshay@kreyerhoff.de} and Georg Kreyerhoff\footnote{georg@kreyerhoff.de}\\
III.~Physikalisches Institut A\\
RWTH Aachen\\
D-52056 Aachen}
\maketitle
\begin{abstract}
We consider a small, metastable maximum vacuum expectation value $b_0$ of order of a few eV, for a
pseudoscalar Goldstone-like field, which is related to the scalar inflaton field $\phi$ in an
idealized model of a cosmological, spontaneously-broken chiral symmetry. The $b$ field allows for relating
semi-quantitatively three distinct quantities in a cosmological context.
\begin{itemize}
\item[(1)] A very small, residual vacuum energy density or effective cosmological constant
of $\sim \lambda b_0^4 \sim 2.7 \times 10^ {-47}\GeV^4$, for $\lambda \sim 3\times 10^{-14}$, the same
as an empirical inflaton self-coupling.
\item[(2)] A tiny neutrino mass, less then $b_0$.
\item[(3)] A possible small variation downward of the proton to electron mass ratio over
cosmological time. The latter arises from the motion downward of the $b$ field over 
cosmological time, toward a nonzero value. Such behavior is consistent
with an equation of motion.
\end{itemize}
We argue that hypothetical $b$ quanta, potentially  inducing new long-range forces, are absent, because
of negative, effective squared mass in an equation of motion for $b$-field fluctuations. The assumed flatness of a potential maximum involves a small inverse-time parameter $\mu \ll 1/t_0$, where
$t_0$ is the present age of the universe.
\end{abstract}

There exist a number of essential physical quantities which probably have a cosmological
origin, and which may have a time variation over cosmological time scales, that is time scales of the
order of the present age of the universe, or much longer. Such quantities are: 
(1) a small, effective cosmological constant,
(2) very small neutrino mass, and (3) a possible small time variation of the proton to electron
mass ratio $(m_p/m_e)$ \cite{ref1,ref1a}. In this paper, we outline the essential elements of a model
which attempts to relate conceptually, and in a numerically approximate manner, the above
phenomena. The model utilizes parameters which characterize initial inflation, that is a scalar
inflaton field $\phi$ which falls from a potential maximum, at or a little above the Planck mass
$M_{Pl}$, to a minimum denoted here by $\phi_0$, just below $M_{Pl}$. The $\phi$ self-coupling
is characterized by an empirically \cite{ref2}, very small, dimensionless coupling $\lambda \sim
3\times 10^{-14}$, and the time scale is of order $ (1/\sqrt{\lambda}\phi_0) \sim 2.4 \times 10^{-36}\secs$ 
(for $\phi_0 \sim 10^{18}\GeV$).\footnote{As counted from the time of $\phi$ leaving its value
$\simgt M_{Pl}$ at the effective potential maximum.}
The  latter corresponds to an inflaton mass of order 
$\sqrt{\lambda}\phi_0 \sim 2\times 10^{11} \GeV$. We have estimated \cite{ref3,ref3a,ref4,ref4a} that (metastable)
inflaton quanta could constitute a significant contribution to cold dark matter in the present
epoch (extremely massive, essentially non-interacting particles). We have illustrated \cite{ref3,ref3a} 
how a maximum, {\underline{ and a minimum}}, of an effective, $\phi$ potential-energy density, can
arise from certain kinds of radiative corrections to a $\lambda \phi^4$ potential-energy density.
Inflation occurs with the inflaton field at the maximum and during its movement to the minimum.
The cosmological chiral model \cite{ref3,ref3a,ref4,ref4a}, spontaneously broken by $\phi_0$,
contains a Goldstone-like \cite{ref5} pseudoscalar
field $b$, so that the self-coupling is $\lambda (\phi^2 + b^2)^2$. In addition, there is a massive,
primordial neutrino-like neutral lepton $L$, with Lagrangian couplings
\be
-g_L (\ol{L} L \phi + i \ol{L}\gamma_5 L b)
\ee
These terms and the self-coupling are initially invariant under the chiral transformations \cite{ref6} (for
infinitesimal $\beta$)
\ba
\phi \to \phi - \beta b, && b\to b + \beta \phi \nn\\
L \to L - \frac{i\beta}{2} \gamma_5 L, && \ol{L} \to \ol{L} -  \frac{i\beta}{2}  \ol{L} \gamma_5
\ea
An essential relationship is contained in  the presence of the $b$ self-coupling $\lambda b^4$, with
the same coupling parameter as for the inflaton. This gives a residual \footnote{ Residual means
after normalizing the minimum in the potential-energy density of the inflaton field to zero.  }
vacuum energy density of $\lambda b_0^4 \sim 2.7 \times 10^{-47} \GeV^4$, if the $b$ field
has a nonzero vacuum expectation value $b_0 \sim 5.5\eV$. A nonzero vacuum expectation value for
a pseudoscalar field, spontaneously breaks CP invariance \cite{ref7,ref7a}, in a cosmological context
\footnote{Spontaneous CP violation is a motivation for considering a nonzero vacuum expectation
value for the $b$ field. For a brief time near to $\sim 10^{-36}\secs$, there can be a CP-violating
effect such as an antineutrino-neutrino asymmetry from a primary, radiation-producing decay process.\cite{ref4,ref4a}  }. 
An attempt is made to make an independent estimate of $b_0$ \cite{ref4,ref4a}, by coupling
the $b$ field to a neutrino, with Lagrangian terms
\be
-{\cal L}_\nu = \tilde{m}_\nu \ol{\nu}\nu + i g_\nu \ol{\nu} \gamma_5 \nu b 
\ee
With the unitary transformation
\be
\nu \to e^{-i\gamma_5 \alpha_\nu/2} \nu,\;\;\; \ol{\nu} \to \ol{\nu} e^{-i\gamma_5 \alpha_\nu/2} 
\ee
for $\tan \alpha_\nu = g_\nu b_0 / \tilde{m}_\nu$, $- {\cal L}_\nu$ becomes \footnote{CP noninvariance
is formally manifest in the scalar coupling of $(\delta b)$, with reference to the pseudoscalar
coupling in (1) (and in (5)).}
\be
-{\cal L}_\nu = m_\nu \ol{\nu} \nu + g_\nu (i \cos \alpha_\nu \ol{\nu}\gamma_5 \nu + \sin\alpha_\nu \ol{\nu}\nu)
(\delta b)
\ee
with $b\to b_0 + \delta b$, $m_\nu = \sqrt{ \tilde{m}_\nu^2 + (g_\nu b_0)^2}$, $\cos \alpha_\nu = \tilde{m}_\nu/m_\nu$,
$\sin\alpha_\nu = g_\nu b_0/m_\nu$. Consider the situation if $\tilde{m}_\nu \ll (g_\nu b_0) $ (essentially,
vanishing ``bare'' neutrino mass $\tilde{m}_\nu$). Then, $\cos \alpha_\nu \sim 0$, $\sin \alpha_\nu \sim 1$;
the effective mass of the neutrino is $\sim (g_\nu b_0)$, numerically $\sim 0.05 \eV$ for \footnote{ In the
numerical, renormalization-group calculations \cite{ref3,ref3a} which give the inflaton potential a maximum
and a minimum near to the Planck mass, we found a representative coupling parameter of $\phi$ (and
$b$) to a primordial, neutrino-like massive lepton $L$, to be $g_L\sim 0.01$. } 
$g_\nu \sim 0.01$. Thus, an explicit breaking of the initial chiral symmetry which is associated with
neutrino mass in Eqs.~(3,5), is of the order of the breaking arising from the spontaneously-broken
CP invariance ($b_0 \neq 0$). The third of our hypothetical connections, concerns an empirically possible time variation,
in particular downward \cite{ref1,ref1a}, of $(m_p/m_e)$. In analogy, with the above argument for neutrino mass,
a hypothetical coupling of the $b$ field to primordial, ordinary quarks gives a quark mass contribution
of $g_b b_0$; subsequently for three confined valence quarks, a nucleon mass contribution of order
$3 g_b b_0$. Here, we have assumed that primordial quarks have zero intrinsic (``bare'') mass. We assume
that electroweak symmetry-breaking generates the standard-model MeV mass contribution for light quarks at
a later time, and we assume that this contribution simply adds to the primordial mass contribution 
estimated here, $g_b b_0$. This is the assumption that the electroweak mass term arises from the Higgs
vacuum expectation value times a tiny coupling to the quark field which has acquired a small mass
term $g_b b_0$. 

Here, we concentrate upon giving an illustrative argument which suggests the consistency of the possibility
for the above third connection. The essential physical assumption is that the $b$ field can fall over
cosmological time intervals, from a maximum value $b_0$, to a nonzero value, say $\sqrt{\epsilon} b_0$ ($
0 < \epsilon < 1$). An effective vacuum energy density is constructed (below) for $b$, which takes as a constant
term $\lambda b_0^4$ with $\lambda$ the tiny self-coupling of the inflaton field. This is suggested by
the cosmological chiral model outlined in the introduction. In our example below, we will consider
$\epsilon$ to be a little less than unity. If the dimensionless parameter $\epsilon$ is not identically
0 ( as usually is assumed ), a value near unity would be, a priori, reasonable (note also the relevance
of $\epsilon \simlt 1$ to the discussion in appendix III).
A second assumption is that potentially new, long-range forces
due to exchange of $b$ quanta do not occur, because of negative, effective squared mass in an equation
of motion for $b$-field fluctuations. It is assumed that quanta with negative squared mass are not present
(i.~e.~superluminal tachyons \cite{ref8,ref8a,ref9,ref9a}). 

A usual equation of motion is
\be
\ddot b+ 3 H \dot b + V'(b) = 0,\;\;\; V'(b) = dV(b)/db
\ee
with $V'(b)$ the derivative of an effective vacuum energy density.
For simplicity in the present argument, we approximate the time-dependent Hubble parameter over long
cosmological times, approximating the present epoch, as functionally,
\be
H(t) \sim \frac{2}{3t} f(\Omega_{vac}(t_0)) \sim 1/t
\ee
for $f(\Omega_{vac}(t_0)) \sim 1.5$, where $ f(\Omega_{vac}(t)) = \Omega^{-1/2}_{vac}  \ln( (1+\Omega^{1/2}_{vac})
/(1-\Omega_{vac})^{1/2})$, $\Omega_{vac} (t) = \rho_{vac} /\rho_C(t)$. The vacuum and critical energy 
densities are $\rho_{vac}$ and $\rho_C(t)$ respectively, with $\Omega_{vac}(t_0)$ assumed to be
$\sim 0.7$ at the age $t_0$ of the present epoch. The factor $f>1$ reflects the fact that the present
age is somewhat greater than $\sim 2/3 H(t_0)$, because of the assumed presence of $\Omega_{vac}$ in an
assumed flat universe \cite{ref10}. First, consider that $V'(b)$ can be neglected in the equation of motion (6).
We then have approximately
\be
\ddot b + 3 \frac{\dot b}{t} = 0
\ee
For this equation, an approximate solution for long cosmological times, greater than some scale $\ol{t}$ which
is significantly less than $t_0$, is
\be
b(t) \sim constant + \frac{(1-\sqrt{\epsilon})b_0}{2} \frac{1}{(t/\ol{t})^2}
\ee
as is verified by differentiation in (8).\footnote{ The essence of the physical argument is not changed
if the fall at large times is stronger, say exponential.} For epochs less than $\ol{t}$, $b$ is assumed to be
nearly constant at a value $b_0$. Now, consider inclusion of a relatively small $V'(b)$ given, as an
explicit example, by the following form which is assumed to hold in the domain $\sqrt{\epsilon}b_0 \le
b \le b_0$, with $\epsilon$ a little less than unity. This involves a small (through $\mu^2$), explicit chiral-symmetry breaking.
\be
V'(b) = \mu^2 \left[ b_0 - \frac{b_0^2}{b} \left\{ \frac{(b^2 - \epsilon b_0^2)}{(1-\epsilon) b_0^2}\right\}^2\right]
\ee
Here, $\mu$ is a fixed parameter with dimension of inverse time, assumed to be set by a time scale
considerably greater than the present epoch $t_0$. Therefore, the $V'$ term in (6) is relatively small, at least out
to times much greater than $t_0$. In (10), we have $V'(b_0) = 0$,
and $V'(b) > 0$ as $b \to \sqrt{\epsilon} b_0$. From (12) below, the latter value corresponds to the zero of $V''(b)$,
i.~e.~the point of change from negative to positive values. We do not extrapolate the form in (10) for $b$
below this point. In particular with $\epsilon$ near zero, such extrapolation of the form (10) has a zero
near $b =\epsilon^2 b_0$, with $V'/\mu^2 b_0$ becoming strongly negative at still lower $b$, and with
$V''/\mu^2$ ( (12) below ) strongly positive. The physical assumption which underlies the speculation in this
paper is that a $b < \sqrt{\epsilon} b_0$ where a $V'(b)$ is zero, is reached only at times much greater
than $t_0$ (and thus than the $\ol{t}$ in (9) ), i.~e.~in epochs characterized by $\mu^{-1}$.
The usual, effective equation of
motion for free propagation of potential quanta of the $b$ field, denoted by $\delta b$, is
\be
\Box (\delta b) + V''(b)(\delta b) = 0
\ee
where $V'' = dV'/db$ gives the effective, squared mass. Evaluating this from (10), gives
\be
V''(b) = \mu^2 \left[ \frac{b_0^2}{b^2} \left\{ \frac{(b^2 - \epsilon b_0^2)}{(1-\epsilon) b_0^2}\right\}^2 
- 4 b_0^2 \frac{ (b^2 - \epsilon b_0^2)}{ ((1-\epsilon) b_0^2)^2} \right] 
\ee
Thus, the effective, squared mass is negative, $\mu^2 (1 - 4/(1-\epsilon))$ at $b=b_0$; it approaches
zero as $b \to \sqrt{\epsilon} b_0$. The positive, very small values of
$V'(b)$ and the negative $V''$, imply that the effective cosmological constant from the energy
density $\cong (\lambda b_0^4 + V'(\Delta b) + V''(\Delta b)^2/2 )$, with $\Delta b < 0$, has a 
completely negligible downward movement. We emphasize that the first, constant
term in this energy density, contains the tiny inflaton self-coupling $\lambda$, as suggested by the
initial chiral symmetry. The energy scale is related to the very small scale of neutrino mass,
of order $b_0$. The extreme flatness (metastability over very long cosmological times) is associated with the
parameter $\mu^2$ in Eqs.~(10,12). This is an essential assumption, which motivates the title of this paper.\footnote{
Assumed contributions of zero-point energies of standard quanta to the vacuum energy density are not
considered in this paper. The presence in the model of a long time (length) scale ($1/\sqrt{\lambda} b_0 $, note
appendix I), might allow for the possibility that these are limited by a smaller value, of order $(\sqrt{\lambda} b_0)^4$.}

The example in the previous paragraph is intended only to illustrate the physical possibility of incorporating
(3) together with the relations (1) and (2), as stated in the opening paragraph of this paper. There are
predictions.
\begin{itemize}
\item[(1)] 
The effective cosmological constant has no discernable time variation, i.~e.~the residual
vacuum energy density in the $b$ field is, in effect, a constant.
\item[(2)]
There are, in principle, discernable, small time variations downward, in neutrino mass,
and in $(m_p/m_e)$ \footnote{ A large coupling parameter $g_b$, of $b$ to quarks, is necessary to reach
the suggested \cite{ref1,ref1a} time variation of ($m_p/m_e$), if attributed to a $\Delta m_p$. The electron mass
can change, but if the leptonic coupling is like that estimated for neutrinos, $g_\nu \sim 0.01$, then
the hypothetical downward change in ($m_p/m_e$) is probably controlled by the downward change in $m_p$.
}, as contributions from the $b$ field go down with $b_0 \to
\sqrt{\epsilon} b_0$.
\item[(3)] Cold dark matter may well involve metastable quanta of the scalar inflaton field
$\phi$, that is super-massive particles at a mass scale of a few times $10^{11}\GeV$. Then,
dark matter is intrinsically related to the existence of the scalar field whose vacuum energy
density induces initial inflation. And the related $b$ field is responsible for entrance
at present into a period of inflation, induced by its vacuum energy density. Thus, in this model, there
is an attempt to relate the different phenomena.
\end{itemize}
The new general idea is that there is a small energy scale $b_0$ associated with the early
universe, in addition to  the usual very high energy scales i.~e.~inflaton field $\phi_0$ and
radiation temperature. In the following appendices, we indicate some possibly suggestive, numerical
properties of certain combinations of these scales. 

\section*{Acknowledgment}
We thank the reviewer for help in clarifying these speculations. 
\section*{Appendix I}

At the end of inflation, there are two very high, energy-scale parameters, $\phi
_0\cong 10^{18} \GeV$
and the initial radiation temperature $T_0 \cong 10^{15.5} \GeV$. But there is a
also a
very small, dimensionless parameter $\lambda$, of the order of $10^{-14}$.\footnote
{Heuristically,
this can be seen by equating the maximum energy density in the inflaton field, to
the
initial (maximum) radiation energy density, which appears just after inflation.
Then, $\sim \lambda M_{\mathrm{Pl}}^4 \sim T_0^4$ gives $\lambda \sim 10^{62}\GeV^4
/10^{76}\GeV^4 =
10^{-14}$. }
A hypothetical small energy scale $ b_0 \cong 5.5 \eV$ is obtained from a conjectured relation
$\lambda^2 = b_0/\phi_0 = 5.5 \times 10^{-27}$ ( for $\lambda = 7.4\times 10^{-1
4}$).
There are then two time-scale parameters: $1/\sqrt{\lambda} \phi_0 = 2.4 \times
10^{-36} \secs$
at the end of inflation; and $1/\sqrt{\lambda} b_0 = 4.4 \times 10^{-10} \secs$.
 Over
the latter time scale, the radiation energy scale evolves downwards to a relatively
low scale
\be
T_0\left( \frac{2.4 \times 10^{-36}\secs}{4.4\times 10^{-10}\secs }\right)^{1/2}
 =
h = 234 \GeV
\ee
This scale is almost the empirical Higgs energy scale (the assumed nonzero vacuum
expectation value v) of the standard model. From this viewpoint, it is a time scale
which determines $h$, i.~e.~the temperature at which a determined value for $h$
is metastable. The coupling $g_e$, of $v/\sqrt{2}$
to electrons, is empirically another small number: $g_e^2 \cong 9\times 10^{-12}
$.
Again, a ratio of energy scales almost gives this very small number $g^2 = b_0 /
h
= 23.5 \times 10^{-12}$.
\footnote{A ratio of energy scales involving $b_0$ is close to
$\alpha^2 = 5.3 \times 10^{-5}$, that is $b_0/m_e = 5.5 \eV/0.5 \MeV =
1.1 \times 10^{-5}$.} In summary, the above emphasis on the role of $\lambda$ allows for a suggestive
numerical hierarchy.
\ba
b_0 &=& \lambda^2 \phi_0 \propto\lambda^2 \nn\\
h &=& \lambda \phi_0 (T_0/\phi_0) \propto\lambda\\
g^2 &=& \lambda (\phi_0/T_0) \propto \lambda ,\;\;\mbox{so}\;\;g \propto \sqrt{\lambda}\nn
\ea
The essential parameters are $\phi_0$, near to $M_{Pl}$, and $\lambda^2 $ ( or alternatively $b_0$,
near to neutrino mass). Here, it is
the extreme smallness of $\lambda$ which relates the high energy scale $\phi_0$( or $T_0$)
to the much lower energy scale $h$.

\section*{Appendix II - ``Cosmic coincidence''}
The present ratio of vacuum energy density $\rho_\Lambda$, to approximate dark-matter energy
density $\rho_{\mathrm{d.m.}}$, empirically $\rho_\Lambda/\rho_{\mathrm{d.m.}}\sim 3$, appears to
be a curious accident, in particular when these densities are assumed to originate in
completely different dynamics, and the matter density falls with the expansion of the universe.
Here, we note that the ratio of order unity is not particularly accidental, when the two densities
are considered together as functions of only two quantities: the very large energy scale
$\phi_0$, and the very small dimensionless parameter $\lambda$.
We have\footnote{For $\lambda  \sim 7.4\times 10^{-14}$, Eq.~(15) gives $\rho_\Lambda$ above the
empirical value by only a factor of about 2.5.}
\ba
\rho_\Lambda &=& \lambda b_0^4 = \lambda (\lambda^2 \phi_0)^4 = \lambda^9 \phi_0^4 \\
\rho_{\mathrm{d.m.}} &=& (\rho_{\mathrm{d.m.}})_0 \left\{\left(\frac{10^{-36}\secs}{\ol{t}}\right)^{3/2}
\left(\frac{\ol{t}}{10^{11}\secs}\right)^{3/2}\left(\frac{10^{11}\secs}{4\times10^{17}\secs}\right)^2\right\}\nn\\
&=& \left( \frac{1}{2}f(\lambda)m_\phi^4 \right) \left\{ (\lambda^4 )^{3/2} \times(\sim \lambda^{0.39})\right\}\\
&=& (\lambda^{2.5} \phi_0^4) \left\{ (\lambda^6)\times (\sim \lambda^{0.39})\right\}\nn\\
&\cong& \lambda^{8.89} \phi_0^4\nn
\ea
In (16), $(\rho_{\mathrm{d.m.}})_0$ is the initial energy density for inflatons 
created at $\sim 10^{-36}\secs$.
This density is written in terms of the inflaton mass $m_\phi \cong \sqrt{\lambda} \phi_0$, as
$ m_\phi^2/2 \times (f(\lambda)m_\phi^2)$ with an assumption for the fluctuation scale \cite{ref3,ref3a},
$\lambda < f(\lambda) < 1$.\footnote{ This is consistent with an estimate of the size of $f$ from
production of massive dark matter by a time-varying gravitational field, when $H$ is of order
$m_\phi$.\cite{ref4,ref4a}} As a definite example, we used in (16) the geometric mean
$f(\lambda) =\sqrt{\lambda}$. The terms bracketed as $\{\ldots\}$ give the time evolution to the
present age of the universe, $\sim 4\times 10^{17}\secs$. The first two terms give 
evolution to $\sim 10^{11} \secs$, taken here as an approximate time of matter dominance. 
The second form of the first term on the r.~h.~s.~explicitly uses a hypothetical $\lambda$ dependence for setting
the scale $\ol{t}$; $\ol{t}$ is determined by $\{(1/m_\phi)\times \ol{t}\}^{1/2} \sim 1/\sqrt{\lambda}b_0 \rightarrow
\ol{t} \sim 10^{-36}\secs / \lambda^4 \sim 3\times 10^{16}\secs$, using $b_0 =\lambda^2 \phi_0$, (this $\ol{t}$
is about a characteristic time for galaxy-halo formation from dark matter).  This conjecture equates a mean time
scale - formed from the product of an initial, very short time scale $1/m_\phi \sim 1/\sqrt{\lambda} \phi_0
\sim 10^{-36}\secs$ (in which the field $b_0$ is established), with the long time scale $\ol{t}$, which
characterizes subsequent, downward change in the $b$ field - to an intrinsic time scale associated with the
small $b_0$.
The last term gives evolution to the present. The last two terms  
are expressed as an effective, small power of $\lambda$. Eqs.~(15,16) give$^{F11}$
\be
\rho_\Lambda/\rho_{\mathrm{d.m.}} \sim 1/30
\ee
This differs from the approximate empirical value by only a factor of about 90. From this point
of view, a ratio of order of unity may be explainable in terms of only two primary, dynamical
quantities, $\phi_0$ and $\lambda^2$ (or alternatively, $b_0$).

\section*{Appendix III: Succession of universes}
The idea in this paper lends itself to a concept of an indefinitely large succession
of universes, which are characterized by an ever-decreasing initial, and residual
vacuum energy density; and by an ever-decreasing ``initial'' radiation and matter
energy density (and thus subsequent matter energy density). Consider a universe
(like ours) with an initial vacuum energy density in the inflaton field
$\lambda\phi^4 \sim  \lambda M_{\mathrm{Pl}}^4 $ (which immediately goes to radiation
and matter, mainly inflaton mass $\sim \sqrt{\lambda}\phi_0$), and a
residual vacuum energy density of order $\lambda b_0^4$ ($\lambda^2$ given by $b_0/\phi_0$,
$b_0\neq 0$, $\phi_0\simlt M_{\mathrm{Pl}}$ ). After an indefinitely long time, an
indefinitely large space(-time) volume exists, with nonzero, residual vacuum energy
density. Consider that in this
large volume a ``speck'' occurs with inflaton field $\sim \sqrt{\epsilon} M_{\mathrm{Pl}} \to \sqrt{\epsilon} \phi_0$ ( so 
$\lambda^2$ is the same tiny dimensionless parameter given by the ratio of fields
$\sqrt{\epsilon}b_0/\sqrt{\epsilon} \phi_0$). Assume that in the speck, a residual vacuum
energy density is established at a maximum value $\lambda(\sqrt{\epsilon}b_0)^4$ ($\epsilon \simlt 1$),
corresponding to the initial field value $\sqrt{\epsilon}b_0$. This speck immediately inflates to
a universe with a residual vacuum energy density reduced by a factor of $(\sqrt{\epsilon})^4
= \epsilon^2$ and with radiation and matter energy densities reduced by $\epsilon^2$.
Repeating this argument $N$ times, each time within an immediately preceding large
space-time volume, gives rise to a universe in which initially, both residual vacuum
energy density and radiation-matter energy density tend to zero, like $(\epsilon^2)^N$ i.~e.~an
empty universe speck (embedded in an infinite space-time volume). Consider as a formal
exercise with $\epsilon$ is a little less than unity, how many times $(N_0)$ this would
have had to occur in the past, in order to have produced effective, quartic self-coupling
of the vacuum field values (i.~e.~$\phi_0,b_0$, as in our universe), which is of order
unity. Then $\lambda / (\epsilon^2)^{N_0} \sim 1 \to N_0 \sim 1.6 \times 10^2$ for
an $\epsilon$ of order 0.9. ($\phi_0$ and $b_0$ would have increased by a factor of $10^{3.5}$.
The initial radiation temperature $T_0$, would be of order $M_{\mathrm{Pl}}$. Alternatively,
given an initial $T_0 \sim 10^{15.5}\GeV$ in our universe, in order that an energy density
in an earlier universe not exceed $M_{\mathrm{Pl}}^4$, $\lambda$ must be as small as
about $(10^{-3.5})^4 = 10^{-14}$). Ours arises as one of the ``livable'' universes,
in a succession of embedded universes with the same initial, dimensionless parameters $\lambda$
and $ \epsilon$.\footnote{In a certain sense, a (classical) direction of time is defined
by a positive value of the pseudoscalar field $b_0$, with the conjecture $\lambda^2 = b_0/\phi_0$.
For $t \to -t, b_0 \to -b_0$, but $\lambda^2 \to -\lambda^2$ is not permissible.}

\end{document}